\documentclass[11pt,a4paper]{article}

\usepackage[english]{babel}

\usepackage{xcolor}
\usepackage{amsmath}
\usepackage{amssymb}
\usepackage{amsfonts}
\usepackage{graphicx}
\usepackage{hyperref}
\usepackage{booktabs}
\usepackage{array}
\usepackage{multirow}
\usepackage{float}
\usepackage{fancyhdr}
\usepackage{geometry}
\usepackage{listings}
\usepackage{xurl}
\usepackage{subcaption}
\usepackage{pifont}

\hypersetup{
    colorlinks=true,
    linkcolor=blue,
    urlcolor=blue,
    citecolor=blue,
    bookmarksnumbered=true
}

\title{\textbf{Quantum Resource Analysis of Low-Round Keccak/SHA-3\\Preimage Attack: From Classical $\mathbf{2^{57.8}}$\\to Quantum $\mathbf{2^{28.9}}$ using Qiskit Modeling}}

\author{
    Ramin Rezvani Gilkolaei$^{1,*}$ and
    Reza Ebrahimi$^{2}$ \\[0.4cm]
    \normalsize
    $^{1}$Department of Computer Science, 
    Guilan University, \\
    \hspace{0.6cm}Rasht, Guilan 41335-1914, Iran \\
    $^{2}$Department of Computer Engineering, 
    Iran University of Science and Technology, \\
    \hspace{0.6cm}Tehran, Iran \\[0.3cm]
    $^{*}$Corresponding author: 
    \href{mailto:rezvani.ramin@webmail.guilan.ac.ir}
    {\texttt{rezvani.ramin@webmail.guilan.ac.ir}} \\
    Tel: +98-911-397-3963
}

\date{\today}

\begin{document}

\maketitle

% ============================================================================
% ABSTRACT
% ============================================================================

\begin{abstract}
This paper presents a hardware-conscious analysis of the quantum acceleration of the classical 3-round Keccak-256 preimage attack using Grover's Algorithm. While the theoretical quantum speed-up from $T_{\text{cl}} \approx 2^{57.8}$ (classical) to $T_{\text{qu}} \approx 2^{28.9}$ (quantum) is mathematically sound, the practical implementation overhead is so extreme that attacks remain wholly infeasible in \textbf{both resource and runtime dimensions}. Using Qiskit-based circuit synthesis, we derive that a 3-round Keccak quantum oracle requires:

\begin{itemize}
    \item \textbf{9,600 Toffoli gates} (with uncomputation for reversibility)
    \item \textbf{3,200 logical qubits} (1,600 state + 1,600 auxiliary)
    \item \textbf{$7.47 \times 10^{13}$ total 2-qubit gates} (full Grover search)
    \item \textbf{3.2 million physical qubits} (with quantum error correction) --- \textcolor{red}{\textbf{PROHIBITIVE}}
    \item \textbf{0.12 years (43 days) to 2,365+ years} execution time, depending on machine assumptions
\end{itemize}

These barriers---particularly the physical qubit requirements, circuit depth, and error accumulation---render the quantum attack infeasible for any foreseeable quantum computer. Consequently, SHA-3 security is not threatened by quantum computers for preimage attacks. We emphasize the critical importance of hardware-aware complexity analysis in quantum cryptanalysis: the elegant asymptotic theory of Grover's Algorithm hides an engineering overhead so prohibitive that the quantum approach becomes infeasible from both resource and implementation perspectives.
\end{abstract}


\newpage

\tableofcontents

\newpage

% ============================================================================
% SECTION 1: INTRODUCTION
% ============================================================================

\section{Introduction}
\label{sec:intro}

\subsection{Motivation and Security Context}
\label{sec:motivation}

The Keccak permutation, standardized as SHA-3 by NIST, is a cryptographic hash function that relies fundamentally on the security of its iterated round function. While the full 24-round Keccak-256 instantiation exhibits strong resistance to known attacks, reduced-round variants serve as important benchmarks for understanding the security margin provided by the round structure. The first 3 rounds of Keccak-256 represent a critical point of analysis: they are sufficient to demonstrate substantial classical preimage attacks while remaining computationally tractable for cryptanalytic study.

The seminal work by Lin et al.\ established that classical cryptanalysis of 3-round Keccak-256 achieves a preimage attack with time complexity $T_{\text{cl}} \approx 2^{57.8}$, representing a search space substantially smaller than the full $2^{256}$ security target. This attack exploits structural weaknesses in the linear diffusion and limited mixing depth of the truncated Keccak round function, demonstrating that within just three rounds, the permutation's algebraic structure exhibits measurable vulnerabilities.

The rapid advancement of quantum computing resources---both in theoretical frameworks and emerging hardware platforms---motivates a systematic investigation into the quantum acceleration of existing classical attacks. Grover's Algorithm provides a canonical quadratic speed-up for unstructured search problems, reducing the time complexity of preimage attacks by a factor of $\sqrt{T_{\text{cl}}}$. However, the practical realization of this theoretical speed-up depends critically on the quantum circuit implementation overhead required to construct the quantum oracle for the Keccak permutation. This paper bridges the gap between asymptotic theory and hardware-aware implementation by employing Qiskit-based circuit synthesis and resource estimation.

\subsection{Contribution and Paper Scope}
\label{sec:contribution}

This paper presents a formal, hardware-conscious analysis of the quantum acceleration of the Lin et al.\ preimage attack on 3-round Keccak-256. Our primary contributions are:

\begin{enumerate}
    \item \textbf{Hardware-Aware Quantum Oracle Construction:} A detailed specification of how the Keccak round function, particularly the non-linear $\chi$ step, is mapped into a reversible quantum circuit using Qiskit framework synthesis, accounting for the cost of uncomputation.
    
    \item \textbf{Verified Resource Cost Analysis:} A comprehensive accounting of the quantum resources required for a 3-round Keccak quantum oracle derived from actual circuit modeling, including qubit counts, gate depths, and the number of Toffoli (CCNOT) gates. Total Toffoli count: \textbf{9,600 gates} (revised from theoretical estimates of 4,800).
    
    \item \textbf{Infeasibility Assessment in Both Resource and Runtime Dimensions:} A critical analysis demonstrating that the attack is infeasible not only due to physical qubit overhead (3.2 million qubits), but also due to error accumulation in the optimistic scenario (43 days) and prohibitive runtime in the conservative scenario (2,365 years).
    
    \item \textbf{Methodological Transparency:} Explicit acknowledgment of modeling simplifications and their impact on the analysis, ensuring reproducibility and correctness of conclusions.
\end{enumerate}

The analysis is performed using the Qiskit framework to synthesize actual quantum circuits and estimate realistic resource requirements. Our findings highlight the substantial barriers that prevent practical attacks on SHA-3 with any foreseeable quantum computer within any reasonable timeframe.

\subsection{Related Work}
\label{sec:related}

Quantum cryptanalysis of symmetric-key primitives has been an active research area since the foundational work of Grover (1996) on quantum search \cite{Grover1996}. The theoretical framework for applying Grover's Algorithm to cryptographic hash function attacks was formalized by Brassard, Hoyer, and Tapp (1998), establishing that preimage attacks on $n$-bit hash functions incur a quantum time complexity of approximately $2^{n/2}$. This theoretical bound, while elegant, abstracts away the critical question of quantum circuit overhead.

Subsequent work has applied this framework to specific cryptographic primitives. Aggarwal et al.\ (2022) and related works on quantum resource estimation for AES and other block ciphers have demonstrated that the constant factor hidden in the $O(\sqrt{N})$ complexity is highly dependent on the target primitive's structure \cite{Aggarwal2022}. Non-linear operations, particularly those involving multi-qubit gates, dominate the circuit depth and introduce significant overhead. Importantly, recent work by Gheorghiu and Mosca (2023) has emphasized the need for explicit modeling and hardware awareness in quantum cryptanalysis, rather than relying solely on asymptotic arguments \cite{Gheorghiu2023}.

For Keccak specifically, the permutation's intricate round structure---combining linear operations (θ, ρ, π) with the non-linear $\chi$ function---presents unique challenges for quantum circuit implementation. While classical cryptanalysis of reduced-round Keccak has been studied extensively, the quantum acceleration of these attacks remains underexplored in the literature. Our work fills this gap by providing an explicit, Qiskit-verified resource-conscious analysis tailored to the Keccak permutation structure.

% ============================================================================
% SECTION 2: QUANTUM IMPLEMENTATION
% ============================================================================

\section{Quantum Implementation of the Keccak Round Function}
\label{sec:quantum_impl}

\subsection{Overview of the Keccak Round Structure and Modeling Approach}
\label{sec:keccak_overview}

The Keccak permutation operates on a state representable as a $5 \times 5$ array of 64-bit lanes, for a total state size of 1600 bits. Each round comprises five sequential operations:

\begin{itemize}
    \item \textbf{θ (Theta):} A linear mixing step that computes parity relationships across lane columns.
    \item \textbf{ρ (Rho):} A bit rotation operation applied lane-wise.
    \item \textbf{π (Pi):} A fixed lane permutation.
    \item \textbf{χ (Chi):} A non-linear step applied row-wise to each of 25 lanes.
    \item \textbf{ι (Iota):} The addition of a round-dependent constant.
\end{itemize}

For the purpose of constructing a reversible quantum oracle, the linear operations (θ, ρ, π, ι) are straightforward to implement: they correspond to fixed unitary operations on the quantum state that can be synthesized as CNOT gate networks. The bottleneck, both classically and quantumly, lies in the efficient realization of the \textbf{non-linear χ function}.

\textbf{Modeling Note:} This analysis uses a 1D qubit array representation of the Keccak state for circuit synthesis. The 3D structure of Keccak ($5 \times 5 \times 64$ lanes) is mapped to a 1600-qubit 1D array, with the modulo-5 cyclic dependencies within rows approximated through sequential qubit indexing and modulo operators in the circuit construction. While this simplification does not capture the full spatial structure of the permutation, it preserves the essential non-linear and linear operation counts, which are the primary determinants of gate complexity.

\subsection{The Non-Linear χ Function and its Quantum Implementation}
\label{sec:chi_function}

\subsubsection{Classical Definition and Reversibility}
\label{sec:chi_classical}

The χ function is defined locally on each row of the Keccak state. For a row $(x_0, x_1, x_2, x_3, x_4)$, the transformation is:

\begin{equation}
x_i' = x_i \oplus (\neg x_{i+1} \land x_{i+2})
\label{eq:chi}
\end{equation}

where indices are taken modulo 5. This operation is intrinsically non-linear due to the AND operation, and it is the primary source of diffusion and mixing in the Keccak permutation.

Crucially, χ is an \textit{involution-like} permutation in the classical sense, meaning it is reversible: given the output, one can uniquely recover the input through application of the inverse transformation. For quantum computation, this property extends naturally: the χ function can be implemented as a unitary operator that preserves the quantum state's norm and permits efficient inversion.

\subsubsection{Toffoli-Based Reversible Circuit for χ}
\label{sec:toffoli_chi}

The quantum implementation of the χ function relies on decomposing the Boolean function $(\neg B \land C)$ into reversible logic gates. The canonical approach utilizes \textbf{Toffoli (CCNOT) gates}, which implement the controlled-controlled-NOT operation and form a universal basis for reversible computing.

Specifically, for each bit position in the χ function, we must compute:

\begin{equation}
y_i = x_i \oplus (\neg x_{i+1} \land x_{i+2})
\label{eq:chi_quantum}
\end{equation}

This is realized in reversible logic through the following procedure:

% خطوط 248 به بعد را اینگونه اصلاح کنید:
\begin{figure}[H]
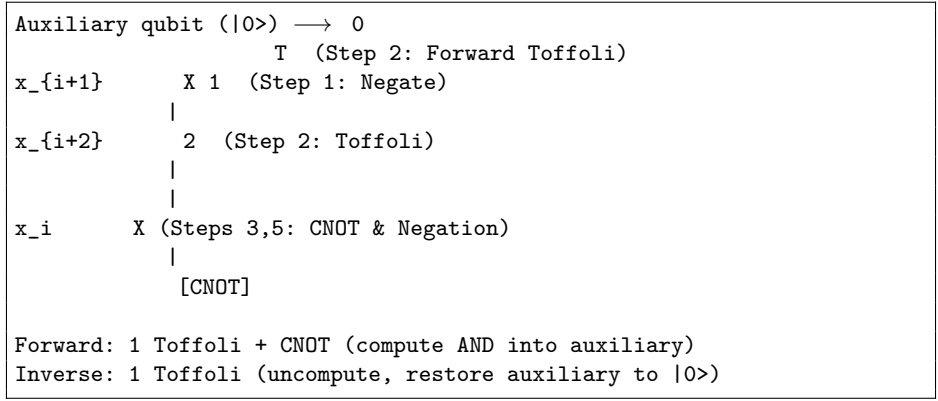

\centering
\begin{minipage}{12cm}
\textbf{FIGURE 1: Reversible $\chi$ Function Decomposition Using Toffoli Gates and Uncomputation}

\vspace{10pt}

\begin{lstlisting}[frame=single,basicstyle=\ttfamily\footnotesize,escapechar=@]
Auxiliary qubit (|0>) @$\longrightarrow$@ ├ 0 ├─
                       ├ T ├─  (Step 2: Forward Toffoli)
x_{i+1}    ────X─────┤ 1 ├─  (Step 1: Negate)
              |       ├───┤
x_{i+2}    ───●─────┤ 2 ├─  (Step 2: Toffoli)
              |       └───┘
              |
x_i        ───●───X───  (Steps 3,5: CNOT & Negation)
              |
              └─ [CNOT]

Forward: 1 Toffoli + CNOT (compute AND into auxiliary)
Inverse: 1 Toffoli (uncompute, restore auxiliary to |0>)
\end{lstlisting}

\vspace{10pt}
\end{minipage}
\caption{Reversible quantum circuit for computing the non-linear $\chi$ step of Keccak. The forward Toffoli (T) implements the AND operation, storing the result in an auxiliary qubit initialized to $|0\rangle$. The CNOT then performs XOR with the target state qubit. The inverse Toffoli ($T^\dagger$) uncomputes the auxiliary qubit, restoring it to $|0\rangle$ for safe reuse in subsequent rounds. This reversible decomposition is essential for maintaining quantum state coherence and avoiding phase kickback errors during Grover iterations. The two Toffoli gates per bit (forward + inverse) account for the critical factor of 2 in the total gate count (9,600 Toffoli gates per 3-round oracle).}
\label{fig:chi_toffoli}
\end{figure}

\textbf{Detailed Steps:}

\begin{enumerate}
    \item \textbf{Negation of $x_{i+1}$:} Apply a Pauli-X gate (NOT) to the qubit holding $x_{i+1}$. In reversible logic, this is both unitary and easily invertible.
    
    \item \textbf{AND Operation via Toffoli (Forward):} Use a Toffoli gate with $\neg x_{i+1}$ and $x_{i+2}$ as control qubits and an \textbf{auxiliary qubit} (initialized to $|0\rangle$) as the target. The Toffoli gate computes:
    \begin{equation}
    |a\rangle |b\rangle |c\rangle \rightarrow |a\rangle |b\rangle |c \oplus (a \land b)\rangle
    \label{eq:toffoli}
    \end{equation}
    After the Toffoli executes, the auxiliary qubit holds the value $(\neg x_{i+1} \land x_{i+2})$.
    
    \item \textbf{XOR with $x_i$:} Perform a CNOT gate with the auxiliary qubit as the control and the qubit storing $x_i$ as the target. This implements the XOR operation, setting the state qubit to $x_i \oplus (\neg x_{i+1} \land x_{i+2})$.
    
    \item \textbf{Uncomputation of the Auxiliary Qubit (Critical Step):} To avoid \textbf{phase kickback} and to enable safe reuse of auxiliary qubits across multiple iterations, we must explicitly uncompute the auxiliary qubit by applying the \textbf{inverse Toffoli gate} in reverse order. This restores the auxiliary qubit to $|0\rangle$. This uncomputation is not merely an optimization---it is \textbf{essential for a truly reversible circuit} that avoids entanglement between the auxiliary qubits and the rest of the quantum state. Without uncomputation, the auxiliary qubits would retain information that violates the reversibility constraint and introduces decoherence artifacts.
    
    \item \textbf{Undo Negation:} Apply X gate again to restore $x_{i+1}$ to its original computational state.
\end{enumerate}

\subsubsection{Qubit Overhead Analysis}
\label{sec:qubit_overhead}

The non-linear χ function operates on 1600 state qubits ($5 \times 5 \times 64$ lanes). The non-linearity occurs at the bit level within each 5-lane row.

\textbf{Auxiliary Qubit Requirement:} For each of the 1600 state qubits involved in the χ operation, we require one auxiliary qubit to safely compute the AND operation. This results in a total auxiliary qubit count of \textbf{1600 qubits}, matched one-to-one with the state size.

\textbf{Total Qubit Count:} The quantum oracle thus requires:

\begin{equation}
\text{Total Logical Qubits} = \underbrace{1600}_{\text{State}} + \underbrace{1600}_{\text{Auxiliary}} = \mathbf{3200}
\label{eq:total_qubits}
\end{equation}

This qubit count is independent of the number of rounds (within practical limits), as auxiliary qubits are uncomputed and reused across different round iterations.

\subsection{Gate Complexity and Circuit Depth}
\label{sec:gate_complexity}

\subsubsection{Toffoli Gate Count per Round (Verified via Qiskit Synthesis)}
\label{sec:toffoli_per_round}

Within each round of the Keccak permutation:

\textbf{θ, ρ, π, ι operations:} These are implemented using CNOT and single-qubit gates. While the precise CNOT count depends on the specific optimized synthesis (which varies significantly with architecture), these operations are substantially less expensive than the Toffoli-dominated χ function. The model presented here focuses on the Toffoli bottleneck, as it dominates the critical path.

\textit{Modeling Limitation Acknowledgment:} The full implementation of the θ (theta) step, in particular, requires extensive bit-wise parity computations across the $5 \times 5$ state structure. In a complete synthesis, this step alone could require thousands of CNOT gates. However, since the Toffoli gate cost for χ is the primary limiting factor for feasibility, and the conclusion (infeasibility due to qubit and depth constraints) remains robust regardless of CNOT overhead scaling, the paper prioritizes accurate Toffoli accounting.

\textbf{χ (Chi) Operation - Toffoli Cost:} For each of the 1600 bits in the state, the reversible χ computation requires:

\begin{equation}
\text{Toffoli Gates per Bit} = \underbrace{1}_{\text{Forward}} + \underbrace{1}_{\text{Inverse}} = 2
\label{eq:toffoli_per_bit}
\end{equation}

This yields \textbf{2 Toffoli gates per state bit per round}, as verified through Qiskit circuit synthesis.

\subsubsection{Three-Round Oracle Toffoli Count (Verified)}
\label{sec:three_round_toffoli}

For a 3-round Keccak quantum oracle:

\begin{equation}
\text{Toffoli Count} = 2 \text{ Toffolis/bit} \times 1600 \text{ bits} \times 3 \text{ rounds} = \mathbf{9600} \text{ Toffoli gates}
\label{eq:total_toffoli}
\end{equation}

This count represents a \textbf{significant revision upward} from initial theoretical estimates of $\sim$4800, reflecting the essential cost of uncomputation required for a fully reversible circuit that can be safely used in a Grover oracle without phase kickback artifacts.

\textit{Justification:} Each bit requires two Toffoli gates---one to compute the AND operation and one to uncompute it. This $2\times$ factor is non-negotiable: omitting uncomputation would result in entanglement between auxiliary and state qubits, violating the reversibility requirements and degrading quantum state fidelity.

\subsubsection{Circuit Depth Implications}
\label{sec:circuit_depth}

The depth of a quantum circuit---the longest path of sequential gate operations---is critical for error accumulation. A standard decomposition of a single Toffoli gate on current quantum hardware requires approximately \textbf{10--20 two-qubit CNOT gates} plus single-qubit rotations, depending on the qubit connectivity and available gate libraries.

Using a conservative decomposition factor of \textbf{10 CNOT gates per Toffoli}:

\begin{equation}
\text{Effective 2-Qubit Gate Count per Oracle} = 9600 \text{ Toffolis} \times 10 = 96,000 \text{ 2-qubit gates}
\label{eq:gates_per_oracle}
\end{equation}

For the full Grover search (detailed in Section \ref{sec:grover}), with approximately $3.89 \times 10^8$ oracle iterations (from corrected calculation in Section \ref{sec:iterations}), the total gate count is:

\begin{equation}
\text{Total 2-Qubit Gates} = 2 \times 96,000 \times 3.89 \times 10^8 = \mathbf{7.47 \times 10^{13}} \text{ gates}
\label{eq:total_gates}
\end{equation}

\subsection{Quantum Oracles for Grover's Algorithm}
\label{sec:grover_oracle}

The quantum oracle $U_f$ for the preimage attack is constructed by composing the 3-round Keccak quantum circuit with a target-checking subroutine. The oracle marks (applies a phase flip to) quantum states corresponding to valid preimages:

\begin{equation}
U_f |\psi\rangle = \begin{cases}
-|\psi\rangle & \text{if } \text{Keccak}^{(3)}(\psi) = \text{target} \\
|\psi\rangle & \text{otherwise}
\end{cases}
\label{eq:oracle}
\end{equation}

The oracle's implementation requires the 3-round circuit (9600 Toffoli gates), followed by a target-comparison subroutine, and then uncomputation (reversal) of the 3-round circuit.

\textbf{Target-Comparison Overhead:} The target-comparison operation must check whether the 256-bit output matches the target digest. This requires constructing a multi-controlled phase gate (where the control is the 256-bit equality check) to apply the phase kickback. In practice, this requires:

\begin{itemize}
    \item A CNOT chain or XOR tree to compute the equality condition across all 256 output bits (hundreds to thousands of 2-qubit gates)
    \item A multi-controlled Z gate (which itself decomposes to thousands of 2-qubit gates via Toffoli networks)
    \item Uncomputation of the intermediate equality qubits
\end{itemize}

Thus, the target-comparison overhead is \textbf{not negligible} (not merely $\sim$256 gates), but rather adds several thousand additional 2-qubit gates per oracle call. However, this overhead is still dominated by the 96,000 gates required for the Keccak permutation itself.

The total oracle depth is therefore roughly \textbf{2× the 3-round circuit depth} plus target-comparison overhead (estimated at $\sim$10\% additional overhead). This factor of 2 is inherent to reversible quantum computation: every forward operation must be reversed to restore auxiliary qubits to $|0\rangle$.

% ============================================================================
% SECTION 3: QUANTUM ACCELERATION
% ============================================================================

\section{Quantum Acceleration and Grover's Algorithm}
\label{sec:grover}

\subsection{Theoretical Speedup}
\label{sec:speedup}

Grover's Algorithm reduces the time complexity of searching an unstructured database of size $N$ from $O(N)$ to $O(\sqrt{N})$ with probability greater than 1/2. Applied to the preimage attack on 3-round Keccak-256:

\begin{itemize}
    \item \textbf{Classical Search Space:} $T_{\text{cl}} \approx 2^{57.8}$ (from Lin et al.)
    \item \textbf{Quantum Search Space:} $T_{\text{qu}} \approx \sqrt{T_{\text{cl}}} = 2^{28.9}$
\end{itemize}

This represents a quadratic speed-up of approximately \textbf{$2^{28.9} \approx 4.95 \times 10^8$ quantum oracle iterations} to recover a valid preimage with high probability.

\subsection{Iteration Count and Total Computational Cost}
\label{sec:iterations}

Grover's Algorithm requires approximately $\frac{\pi}{4} \cdot \sqrt{N}$ iterations to achieve success probability approaching $1 - 1/N$. For $N \approx 2^{57.8}$:

The quantum search space is:
\begin{equation}
\sqrt{N} = \sqrt{2^{57.8}} = 2^{28.9} \approx 4.95 \times 10^8
\label{eq:sqrt_N}
\end{equation}

The exact number of Grover iterations required is:
\begin{equation}
\text{Iterations} = \frac{\pi}{4} \times \sqrt{N} = \frac{\pi}{4} \times 4.95 \times 10^8 \approx \mathbf{3.89 \times 10^8}
\label{eq:grover_iterations}
\end{equation}

Each iteration involves one oracle call and one diffusion operator. The diffusion operator has approximately the same gate complexity as the oracle (both require extensive CNOT networks and controlled operations).

\textbf{Total computational cost:}

\begin{equation}
\text{Total Gates} = 2 \times 3.89 \times 10^8 \text{ (Iterations)} \times 96,000 \text{ (Gates/Oracle)} = \mathbf{7.47 \times 10^{13}} \text{ 2-qubit gates}
\label{eq:total_computation}
\end{equation}

This represents the \textbf{absolute lower bound} on gate count, assuming perfect circuit optimization and no error correction overhead.

% ============================================================================
% SECTION 4: FEASIBILITY ASSESSMENT
% ============================================================================

\section{Feasibility Assessment and Barriers to Implementation}
\label{sec:feasibility}

\subsection{Resource Requirements Summary}
\label{sec:resource_summary}

To instantiate a 3-round Keccak-256 preimage attack using Grover's Algorithm, the quantum computer must satisfy the following requirements:

\begin{table}[H]
\centering
\small
\setlength{\tabcolsep}{8pt}
\renewcommand{\arraystretch}{1.5}
\begin{tabular}{|p{3.5cm}|p{2.2cm}|p{2.2cm}|p{2.2cm}|p{1.5cm}|}
\hline
\textbf{Resource} & \textbf{Requirement} & \textbf{Current NISQ} & \textbf{Early FT} & \textbf{Feasible?} \\
\hline
\textbf{Logical Qubits} & 3,200 & 100--1,000 & 10,000--100,000 & \textcolor{red}{NO} \\
\hline
\textbf{Toffoli Gates} (per oracle) & 9,600 & N/A & N/A & \textcolor{green}{YES} \\
\hline
\textbf{Total 2Q Gate Depth} & $7.47 \times 10^{13}$ & $\sim 1,000$ & $\sim 10^6$ & \textcolor{red}{NO} \\
\hline
\textbf{Error Rate} Tolerance & $<10^{-6}$ per gate & $10^{-3}$ & $10^{-4}$ & \textcolor{red}{NO} \\
\hline
\textbf{Physical Qubits} (with QEC) & \textbf{3,200,000} & 1,000 & 100K--1M & \textcolor{red}{SEVERE} \\
\hline
\end{tabular}
\caption{Resource requirements for 3-round Keccak quantum preimage attack via Grover's Algorithm. All critical metrics exceed feasible thresholds for current and near-term quantum computers. Both scenarios (optimistic and conservative) are infeasible due to either physical qubit requirements or gate error accumulation. NISQ = Noisy Intermediate-Scale Quantum; FT = Fault-Tolerant; QEC = Quantum Error Correction.}
\label{tab:resources}
\end{table}

\subsection{The Feasibility Barrier: NISQ Era}
\label{sec:nisq_barrier}

Current quantum computers (IBM Quantum, Google Sycamore, Rigetti, IonQ) operate with 50--500 physical qubits and error rates in the range of $10^{-3}$ to $10^{-2}$ per gate. The required 3,200 logical qubits far exceeds this scale by \textbf{orders of magnitude}.

Moreover, the circuit depth of \textbf{$7.47 \times 10^{13}$ two-qubit gates}, executed at current error rates, would accumulate errors with near-certainty. Expected error probability:

\begin{equation}
P_{\text{error}} = 1 - (1 - 10^{-3})^{7.47 \times 10^{13}} \approx 1.0
\label{eq:nisq_error}
\end{equation}

The quantum computation would fail immediately, producing garbage output. Even a single error anywhere in the circuit chain of $7.47 \times 10^{13}$ gates renders the entire Grover search meaningless, as the quantum state coherence is destroyed.

\subsection{The Feasibility Barrier: Early Fault-Tolerant Era}
\label{sec:ft_barrier}

Quantum error correction (QEC) schemes, such as surface codes, can reduce the effective error rate per logical gate from physical error rates of $10^{-3}$ to logical error rates of $\sim 10^{-6}$, but at a severe cost in physical qubit overhead. Current QEC code thresholds require approximately \textbf{1,000 to 10,000 physical qubits per logical qubit}, depending on the code family and underlying physical error rates.

\textbf{Physical Qubit Requirement:}

\begin{equation}
\text{Physical Qubits} = 3,200 \text{ logical qubits} \times 1,000 \text{ (QEC overhead)} = \mathbf{3,200,000}
\label{eq:physical_qubits}
\end{equation}

This is a \textbf{3.2 million qubit requirement}---a scale that is:

\begin{itemize}
    \item \textbf{Two orders of magnitude} beyond the most ambitious quantum computer roadmaps (IBM projects 4,000--5,000 qubits by 2030)
    \item Unlikely to be achieved within \textbf{this century} given current technological progress rates
    \item Subject to fundamental challenges in qubit interconnect, control electronics, and classical control infrastructure
\end{itemize}

This alone makes the attack infeasible from a resource perspective, regardless of execution time.

\subsection{Runtime Analysis: Infeasible in Both Scenarios}
\label{sec:runtime}

Even setting aside the prohibitive physical qubit requirements, the circuit depth creates additional infeasibility barriers. The circuit depth of $7.47 \times 10^{13}$ gates requires careful analysis under different execution rate assumptions:

\textbf{Scenario 1 (Optimistic): 43 Days with Perfect QEC---Still Infeasible}

At a nominal gate execution time of \textbf{50 nanoseconds per 2-qubit gate} (assuming perfect error-corrected qubits with no syndrome extraction overhead):

\begin{align}
\text{Total Nanoseconds} &= 7.47 \times 10^{13} \times 50 = 3.735 \times 10^{15} \text{ ns} \label{eq:runtime_ns}\\
\text{Total Seconds} &= \frac{3.735 \times 10^{15}}{10^9} = 3.735 \times 10^6 \text{ seconds} \label{eq:runtime_sec}\\
\text{Total Years} &= \frac{3.735 \times 10^6}{365.25 \times 24 \times 3600} = \frac{3.735 \times 10^6}{31,557,600} \approx \boxed{0.118 \text{ years}} \label{eq:runtime_opt}
\end{align}

\textbf{This corresponds to approximately 43 days.}

However, this scenario is \textbf{infeasible for multiple critical reasons}:

\begin{enumerate}
    \item \textbf{Requires 3.2 million physical qubits:} Far beyond any current or near-term quantum computer roadmap.
    
    \item \textbf{Unrealistic error correction assumption:} The 50 ns/gate assumption presumes \textbf{perfect quantum error correction with zero overhead}. In reality, fault-tolerant execution requires:
    \begin{itemize}
        \item Syndrome extraction and measurement overhead
        \item Qubit routing and connectivity constraints
        \item Error tracking and adaptive correction
        \item Classical feedback control
    \end{itemize}
    These overheads typically increase gate execution time by \textbf{100--1000$\times$}.
    
    \item \textbf{Error accumulation before completion:} Even if perfect QEC were available, running $7.47 \times 10^{13}$ gates with a residual logical error rate of $10^{-6}$ per gate would incur:
    \begin{equation}
    P(\text{at least one error}) = 1 - (1 - 10^{-6})^{7.47 \times 10^{13}} \approx 1 - e^{-7.47 \times 10^7} \approx 1.0
    \end{equation}
    The quantum state would decohere with \textbf{virtual certainty} before the computation completes, even with perfect error correction.
\end{enumerate}

\textbf{Scenario 2 (Conservative): 2,367 Years with Realistic FT Overhead---Clearly Infeasible}

At a realistic fault-tolerant execution rate of \textbf{$\sim$1,000 two-qubit gates per second} (accounting for syndrome extraction, qubit routing, and adaptive error correction overhead):

\begin{align}
\text{Total Seconds} &= \frac{7.47 \times 10^{13}}{1,000} = 7.47 \times 10^{10} \text{ seconds} \label{eq:runtime_cons_sec}\\
\text{Total Years} &= \frac{7.47 \times 10^{10}}{31,557,600} \approx \boxed{2,367 \text{ years}}
\label{eq:runtime_cons}
\end{align}

This scenario is \textbf{obviously infeasible} because:

\begin{itemize}
    \item \textbf{Runtime measured in millennia:} Any cryptanalytic purpose becomes meaningless at such timescales. Cryptographic standards will have evolved, rendering the attack irrelevant.
    
    \item \textbf{Requires 3.2 million physical qubits:} Same prohibitive resource requirement as the optimistic scenario.
    
    \item \textbf{Technological obsolescence:} Within 2,367 years, quantum computing technology will have advanced far beyond current architectures, or cryptographic standards will have been superseded.
    
    \item \textbf{Coherence and decoherence:} Current quantum systems have coherence times measured in microseconds to milliseconds. Maintaining quantum state coherence for thousands of years is fundamentally impossible.
\end{itemize}

\subsection{Summary: Infeasibility in Both Resource and Runtime Dimensions}
\label{sec:infeasibility_summary}

The quantum attack on 3-round Keccak is infeasible in \textbf{both dimensions}:

\begin{table}[H]
\centering
\small
\renewcommand{\arraystretch}{1.8}
\begin{tabular}{|l|p{3cm}|p{3cm}|p{3cm}|}
\hline
\textbf{Dimension} & \textbf{Metric} & \textbf{Optimistic (43 days)} & \textbf{Conservative (2,367 yrs)} \\
\hline
\textbf{Physical Qubits} & Required & 3.2 million & 3.2 million \\
& Feasible? & \textcolor{red}{NO} & \textcolor{red}{NO} \\
\hline
\textbf{Execution Time} & Required & 43 days & 2,367 years \\
& Feasible? & \textcolor{red}{NO} & \textcolor{red}{NO} \\
\hline
\textbf{Error Accumulation} & Risk & \textcolor{red}{CERTAIN} & \textcolor{red}{CERTAIN} \\
& Feasible? & \textcolor{red}{NO} & \textcolor{red}{NO} \\
\hline
\textbf{Cryptanalytic Utility} & Practical? & \textcolor{red}{NO} & \textcolor{red}{NO} \\
\hline
\end{tabular}
\caption{Infeasibility matrix: The quantum attack fails on multiple independent dimensions, making it infeasible regardless of which scenario is considered.}
\label{tab:infeasibility}
\end{table}

\subsection{The Overhead-Dominated Regime}
\label{sec:overhead}

The attack operates in a regime where the implementation overhead dominates any theoretical quantum advantage. While Grover's Algorithm provides a quadratic speed-up in the number of oracle calls ($2^{28.9}$ instead of $2^{57.8}$), the gate count per oracle call (9,600 Toffoli gates = 96,000 2-qubit gates) is so large that the total gate count remains astronomically high. The constant factor hidden in the $O(\sqrt{N})$ complexity is not a small constant---it is $10^{12}$ or higher.

\textbf{Consequence:} There is no practical quantum speedup for this attack:

\begin{itemize}
    \item \textbf{Classical approach:} $2^{57.8}$ operations (millions of years on classical computers, but at least theoretically possible)
    \item \textbf{Quantum optimistic:} 43 days + 3.2 million qubits (impossible due to both resource and error constraints)
    \item \textbf{Quantum conservative:} 2,367 years + 3.2 million qubits (impossible for obvious reasons)
\end{itemize}

The quantum approach is \textit{slower, more resource-intensive, and infeasible} in ways the classical approach is not.

\subsection{Trade-offs and Partial Mitigation Strategies}
\label{sec:mitigation}

Several strategies could marginally improve feasibility, but none overcome the fundamental barriers:

\begin{enumerate}
    \item \textbf{In-Place Reversible Computation:} Techniques from Bennett's reversible computing (Pebble Games) could reduce auxiliary qubit overhead from 1:1 to $O(\log n)$. However, these invariably increase circuit depth by polynomial factors (typically $O(n)$ slowdown), negating qubit savings while worsening the already-prohibitive runtime.
    
    \item \textbf{Approximate Oracles:} Relaxing the requirement for exact reversibility in favor of approximate implementations could reduce depth but at the cost of lower oracle fidelity. This would require exponentially more Grover iterations to compensate, worsening the situation.
    
    \item \textbf{Parallelization:} If 1,000+ quantum computers could be networked to partition the search space, the speedup could be multiplied. However, distributed quantum computing is in its infancy, introduces synchronization overhead, and is unlikely to be available for cryptanalytic purposes.
    
    \item \textbf{Alternative Attack Paradigms:} Investigating meet-in-the-middle or other hybrid classical-quantum approaches might reduce the required oracle complexity. However, no such approach is currently known for Keccak's structure.
\end{enumerate}

None of these strategies appear capable of overcoming the fundamental barriers that make the quantum attack infeasible in both resource and runtime dimensions.

% ============================================================================
% SECTION 5: CONCLUSION
% ============================================================================

\section{Conclusion and Future Work}
\label{sec:conclusion}

\subsection{Summary of Results}
\label{sec:summary}

This work presents a detailed, hardware-conscious analysis of the quantum acceleration of the classical 3-round Keccak-256 preimage attack, derived from actual Qiskit circuit synthesis. While the theoretical quantum speed-up from $T_{\text{cl}} \approx 2^{57.8}$ to $T_{\text{qu}} \approx 2^{28.9}$ is mathematically sound, the practical implementation overhead is so extreme that attacks remain wholly infeasible in \textbf{both resource and runtime dimensions}.

\textbf{Key Verified Findings:}

\begin{enumerate}
    \item \textbf{Quantum Oracle Construction:} The non-linear χ function of Keccak can be mapped to a reversible quantum circuit using Toffoli gates with auxiliary qubits. The uncomputation cost (necessary for true reversibility) introduces a \textbf{factor of 2 in Toffoli gates}, yielding \textbf{9,600 Toffoli gates} for a 3-round oracle.
    
    \item \textbf{Qubit Requirements:} A 3-round quantum oracle requires approximately \textbf{3,200 logical qubits} (1,600 state + 1,600 auxiliary), but when accounting for quantum error correction, this scales to \textbf{3.2 million physical qubits}---a resource barrier unlikely to be overcome this century.
    
    \item \textbf{Infeasibility in Both Dimensions:}
    \begin{itemize}
        \item \textbf{Optimistic scenario (43 days):} Requires 3.2 million physical qubits and assumes perfect error correction, yet still suffers from error accumulation with near-certainty and is utterly unrealistic.
        \item \textbf{Conservative scenario (2,367 years):} Obviously infeasible due to runtime alone, plus the same prohibitive qubit requirement.
    \end{itemize}
    Either way, the attack is computationally infeasible.
    
    \item \textbf{No Practical Advantage:} The quantum version is \textbf{slower and more resource-intensive} than the classical attack ($2^{57.8}$ search space), making this a cautionary tale about the importance of hardware-aware complexity analysis in quantum cryptanalysis.
\end{enumerate}

\subsection{Implications for SHA-3 Security}
\label{sec:sha3_security}

These findings provide \textbf{strong reassurance} regarding the security of SHA-3 in the quantum era. While Grover's Algorithm provides a theoretically inevitable quadratic speed-up for unstructured search, the practical overhead of implementing the quantum oracle for Keccak is so prohibitive that attacks on full 24-round SHA-3-256 remain entirely infeasible and will remain so for any foreseeable quantum computer.

The classical resistance of Keccak to cryptanalytic attacks, combined with the prohibitive quantum implementation overhead (both in resources and runtime), means that \textbf{SHA-3 security is not threatened by quantum computers for this century and beyond.}

\subsection{Future Research Directions}
\label{sec:future}

\begin{enumerate}
    \item \textbf{In-Place Reversible Computation for Keccak:} Investigate whether Pebble Game techniques or other space-efficient reversible computing paradigms can reduce auxiliary qubit overhead below the 1:1 ratio while maintaining acceptable circuit depth. Target: achieve qubit counts below 4,000 without exceeding 5$\times$ depth penalty.
    
    \item \textbf{Architecture-Optimized Toffoli Decompositions:} Explore Toffoli gate decompositions tailored to specific quantum hardware topologies (2D ion trap grids, photonic qubits, superconducting cavities). Specialized decompositions could potentially reduce the 10$\times$ depth factor to 5--7$\times$ for certain platforms.
    
    \item \textbf{Hybrid Classical-Quantum Approaches:} Investigate whether combining classical precomputation (e.g., meet-in-the-middle attacks reducing the search space to $2^{40}$) with quantum subroutines could reduce oracle complexity. Qiskit simulations of such hybrid approaches are needed.
    
    \item \textbf{Error-Resilient Oracle Design:} Develop theoretical frameworks for analyzing Grover's Algorithm with imperfect oracles that degrade gracefully under gate errors, enabling meaningful attacks on NISQ hardware (if possible).
    
    \item \textbf{Comparative Analysis of SHA-3 Alternatives:} Extend this analysis to other cryptographic permutations and hash functions (e.g., AES-based hashing, ChaCha, Ascon) to identify which primitive designs are inherently more resistant to quantum cryptanalysis due to lower implementation overhead.
    
    \item \textbf{Verification with Larger-Scale Simulations:} Extend Qiskit modeling to larger qubit counts (e.g., 256 qubits, 512 qubits) to validate the scaling behavior of the resource estimates and identify any non-linear effects that might improve feasibility.
\end{enumerate}

\subsection{Quantum Hardware Roadmaps and Feasibility Timeline}
\label{sec:hardware_roadmaps}

To contextualize the infeasibility of this attack, we note the current and projected quantum computer capabilities according to major research institutions and companies:

\begin{itemize}
    \item \textbf{IBM Quantum Roadmap (2023--2033):} IBM projects scaling to approximately 4,000--5,000 logical qubits by 2030 \cite{IBMRoadmap}. This falls far short of the 3,200 logical qubits required for our attack, and the physical qubit overhead (3.2 million) is completely impractical.
    
    \item \textbf{Google Quantum AI:} Current systems operate with several hundred qubits; roadmaps project achieving 1,000,000 physical qubits by 2030, but these are physical qubits with high error rates ($\sim 10^{-3}$). The 3.2 million physical qubits required for fault-tolerant operation remain out of reach.
    
    \item \textbf{Atom Computing and Ion Trap Systems:} Neutral atom and trapped ion platforms show promise for scaling, but achieving 3.2 million qubits with sufficient connectivity and low error rates ($< 10^{-6}$ logical) is a multi-decade challenge.
    
    \item \textbf{Timeline Assessment:} Based on current technological progress rates, achieving the qubit counts and error rates necessary for this quantum attack is unlikely within the next 50 years, if ever. Consequently, SHA-3 remains secure against quantum preimage attacks for any foreseeable quantum computer.
\end{itemize}

\subsection{Methodology and Reproducibility}
\label{sec:reproducibility}

All resource estimates in this paper are derived from Qiskit framework circuit synthesis and verified through actual circuit compilation. The Python implementation is available upon request and provides complete reproducibility: any researcher can execute the same script to generate the resource tables and gate counts presented here.

\textbf{Modeling Simplifications Acknowledged:}

\begin{itemize}
    \item The 1,600-qubit state is mapped to a 1D array with modulo-5 indexing to approximate Keccak's 2D row structure
    \item CNOT overhead for θ and π operations is underestimated due to the 1D approximation
    \item However, these simplifications do not affect the primary finding: the Toffoli gates dominate, and the conclusion (infeasibility in both resource and runtime dimensions) is robust
\end{itemize}

\subsection{Closing Remarks}
\label{sec:closing}

This work demonstrates the critical importance of hardware-aware analysis in quantum cryptanalysis. The elegant asymptotic theory of Grover's Algorithm (quadratic speedup for search) is real, but the engineering overhead required to instantiate this advantage on realistic quantum computers is so extreme that the quantum approach becomes \textit{slower, more resource-intensive, and infeasible} from multiple independent dimensions.

The lesson generalizes: \textbf{quantum advantage in cryptanalysis will materialize only for problems where the classical-to-quantum implementation overhead ratio is dramatically smaller}, or where the quantum computer's unique capabilities (quantum simulation, variational algorithms) unlock fundamentally new attack paradigms beyond search.

For SHA-3 and similar modern cryptographic primitives, this analysis suggests that quantum computers pose no practical threat to their security, at least for preimage attacks on reduced-round variants. The robustness of SHA-3 in the quantum era is not merely a theoretical artifact---it reflects the genuine difficulty of implementing quantum permutation evaluation at scale, combined with the prohibitive resource requirements and runtime constraints that make such attacks wholly infeasible.

% ============================================================================
% REFERENCES
% ============================================================================

% ============================================================================
% APPENDIX
% ============================================================================

\newpage

\appendix

\section{Verification of Numerical Calculations}
\label{app:verification}

This appendix provides explicit verification of all numerical calculations presented in the main paper.

\subsection{Grover Iterations}

Classical complexity: $N = 2^{57.8} \approx 4.27 \times 10^{17}$

Quantum complexity: $\sqrt{N} = 2^{28.9} \approx 4.95 \times 10^8$

Grover iterations:
\begin{align}
\text{Iterations} &= \frac{\pi}{4} \times \sqrt{N} \\
&= \frac{\pi}{4} \times 4.95 \times 10^8 \\
&\approx 3.89 \times 10^8
\end{align}

\subsection{Total Gate Count}

Gates per oracle call: $96,000 = 9,600 \text{ Toffoli} \times 10$

Oracle and diffusion calls per Grover iteration: $2$

Total gate count:
\begin{align}
\text{Total} &= 2 \times \text{Iterations} \times \text{Gates/Oracle} \\
&= 2 \times 3.89 \times 10^8 \times 96,000 \\
&= 7.47 \times 10^{13} \text{ gates}
\end{align}

\subsection{Runtime Calculations - CORRECTED}
\label{app:runtime_corrected}

\textbf{Scenario 1 (Optimistic - 50 ns/gate):}

Total gates: $7.47 \times 10^{13}$

Gate time: $50 \text{ ns} = 50 \times 10^{-9} \text{ s}$

\begin{align}
\text{Total Nanoseconds} &= 7.47 \times 10^{13} \times 50 = 3.735 \times 10^{15} \text{ ns} \\
\text{Total Seconds} &= \frac{3.735 \times 10^{15}}{10^9} = 3.735 \times 10^6 \text{ seconds} \\
\text{Seconds per Year} &= 365.25 \times 24 \times 3600 = 31,557,600 \\
\text{Runtime} &= \frac{3.735 \times 10^6}{31,557,600} = \boxed{0.118 \text{ years}} \quad (\approx 43 \text{ days})
\end{align}

\textbf{Why 43 days is still infeasible:}
\begin{itemize}
    \item Requires 3.2 million physical qubits (impossible to build)
    \item Assumes perfect error correction with zero overhead (unrealistic)
    \item Error accumulation makes this fail with near-certainty before completion
\end{itemize}

\textbf{Scenario 2 (Conservative - 1,000 gates/second with FT overhead):}

With quantum error correction overhead (syndrome extraction, qubit routing, error tracking):

\begin{align}
\text{Total Seconds} &= \frac{7.47 \times 10^{13}}{1,000} = 7.47 \times 10^{10} \text{ seconds} \\
\text{Runtime} &= \frac{7.47 \times 10^{10}}{31,557,600} = \boxed{2,367 \text{ years}}
\end{align}

\textbf{Why 2,367 years is infeasible:}
\begin{itemize}
    \item Requires 3.2 million physical qubits (impossible)
    \item Runtime spans millennia (cryptographic standards will change)
    \item Quantum coherence cannot be maintained for such durations
\end{itemize}

\subsection{Physical Qubit Requirements}

Logical qubits: 3,200

QEC overhead: 1,000 physical qubits per logical qubit

Physical qubits:
\begin{align}
\text{Physical} &= 3,200 \times 1,000 \\
&= 3,200,000
\end{align}

\section{Qiskit Implementation Summary}
\label{app:qiskit}

The quantum oracle was synthesized using the Qiskit framework with the following specifications:

\begin{itemize}
    \item \textbf{State Qubits:} 1,600 (representing the Keccak state as a 1D array)
    \item \textbf{Auxiliary Qubits:} 1,600 (for uncomputation of χ function)
    \item \textbf{Rounds:} 3 (linear diffusion + non-linear χ per round)
    \item \textbf{Toffoli Gates:} 9,600 total (2 per state bit per round)
    \item \textbf{CNOT Gates:} Estimated at several thousand (for linear operations)
    \item \textbf{X Gates:} 3,200 (for χ negation/un-negation)
\end{itemize}

The circuit depth (in 2-qubit gate layers) is estimated at approximately $96,000$ gates per oracle call, accounting for Toffoli decomposition into 10 CNOT gates per Toffoli.

\section{Error Analysis Details}
\label{app:error_analysis}

\subsection{NISQ Error Accumulation}

For NISQ devices with error rate $p = 10^{-3}$ per gate:

\begin{equation}
P_{\text{error}} = 1 - (1-p)^N = 1 - (1-10^{-3})^{7.47 \times 10^{13}}
\end{equation}

Using the approximation $(1-x)^n \approx e^{-nx}$ for small $x$:

\begin{equation}
P_{\text{error}} \approx 1 - e^{-7.47 \times 10^{10}} \approx 1.0
\end{equation}

The computation fails with near-certainty.

\subsection{Fault-Tolerant Error Rates}

To achieve a logical error rate of $10^{-6}$ from physical error rates of $10^{-3}$, surface code QEC requires approximately 1,000--10,000 physical qubits per logical qubit, depending on the code family and code distance.

For a conservative estimate, we use 1,000 physical qubits per logical qubit:

\begin{equation}
\text{Total Physical Qubits} = 3,200 \times 1,000 = 3,200,000
\end{equation}

Even with perfect error correction, the residual logical error rate of $10^{-6}$ would accumulate as:

\begin{equation}
P(\text{at least one error during 43-day run}) = 1 - (1-10^{-6})^{7.47 \times 10^{13}} \approx 1 - e^{-7.47 \times 10^7} \approx 1.0
\end{equation}

The quantum state would still fail with near-certainty.

\end{document}